\documentclass{ws-procs10x7}

\begin{document}

\title{Production of vector mesons and measurement of the
hadronic component of photon light-cone wave function at HERA}

\author{N. Coppola\\ on behalf of the H1 and ZEUS Collaborations}

\address{NIKHEF, Kruislaan 409 1009 DB Amsterdam, Netherlands
\\E-mail: coppola@mail.desy.de}

\twocolumn[\maketitle\abstract{ A detailed study  of
vector meson production ($\phi$ and $J/\psi$) in $e^\pm p$ collisions at HERA
with the ZEUS and H1 detector has been performed.  The cross sections
are measured as a function of $Q^2$, $W$ and $t$.  In this
contribution, the results are summarised, compared
to theoretical calculations and the dynamical picture emerging in
perturbative QCD is highlighted.  
The measurement
of the hadronic component of the photon light-cone wave
function in the exclusive production of di-pions, $ep\rightarrow e
\pi^+\pi^- p$ is also
reported.}]

\section{Exclusive vector meson production}\label{sec:vecmesprod}

The exclusive and inelastic production of vector mesons $ep \rightarrow
e V X$, with $V=\phi, J/\psi$ and with $X$ being either the scattered
proton or the remaining hadronic system, has been extensively
investigated at HERA.

In QCD-based models, the vector meson production is viewed as a
sequence of processes: the virtual photon, $\gamma^*$, fluctuates into a
$q\bar{q}$ pair which subsequently interacts with the proton and
eventually forms a meson bound state. 
At high $\gamma^* p$ center-of-mass energies, $W$,
these successive processes are clearly separated
in time (factorisation). The transverse size of the $q\bar{q}$ pair
depends on the photon virtuality $Q^2$ and on the quark masses; for
$Q^2>{\cal O}(10)~\mbox{GeV}^2$ or for
large masses, it is
considerably smaller than the proton size. At these distances the
strong coupling is small and perturbative QCD (pQCD) can be applied.

HERA offers the unique opportunity to study the dependences of these
processes on several different scales: the mass of the meson, $m_V$,
$W$, $Q^2$ and the four-momentum transfer squared at the proton
vertex, $t$. Strong interactions can be investigated in the transition
from the hard to the soft regime, where the confinement of quarks and
gluons occurs.

\begin{figure}
\epsfxsize180pt
\figurebox{120pt}{160pt}{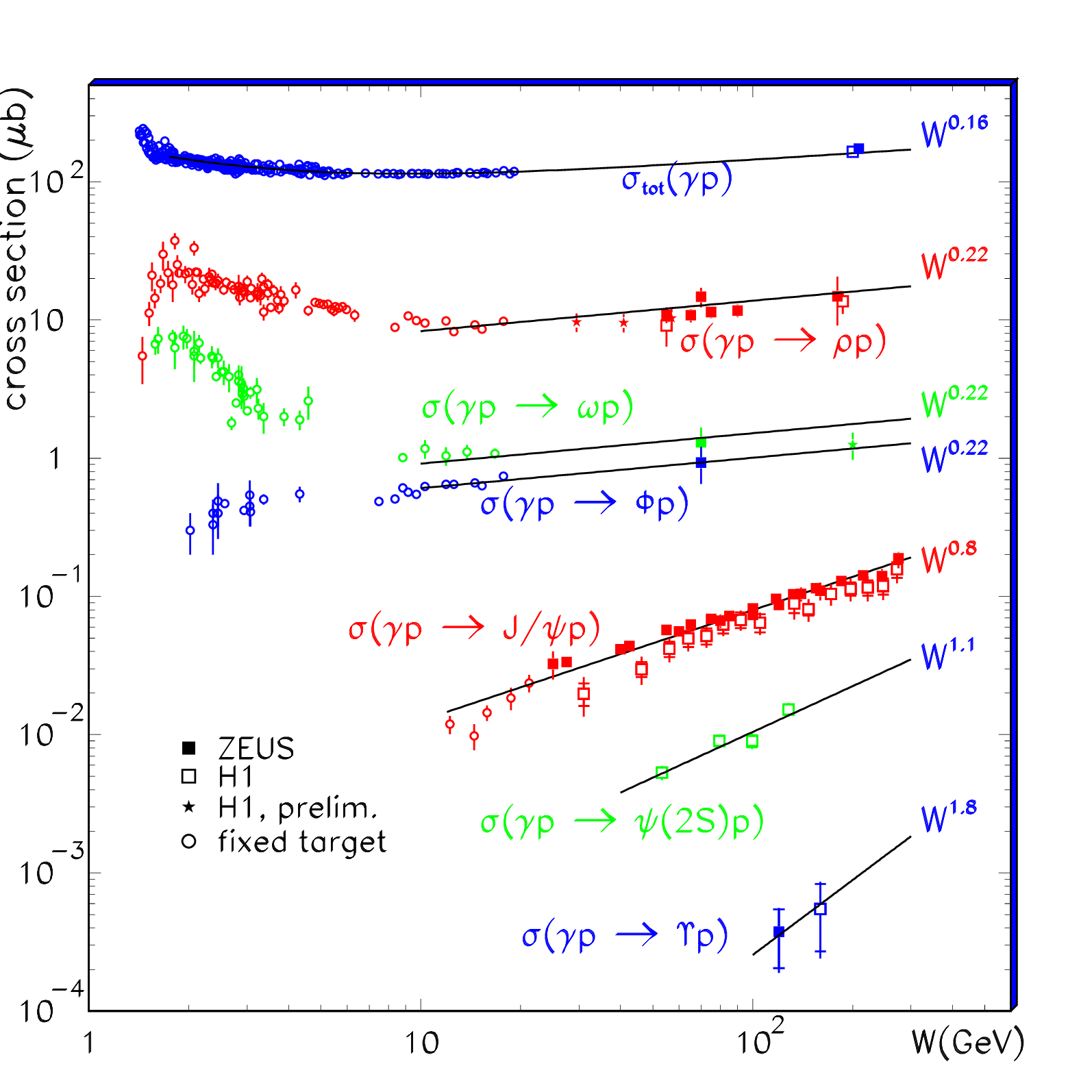}
\caption{Compilation of cross sections for exclusive photoproduction
  of vector mesons as a function of $W$. The lines indicate a power
  law dependence, $W^\delta$. The total photoproduction cross section
  is also shown.
}
\label{fig:wdepxsect}
\end{figure}

\begin{figure}
\epsfxsize180pt
\figurebox{120pt}{160pt}{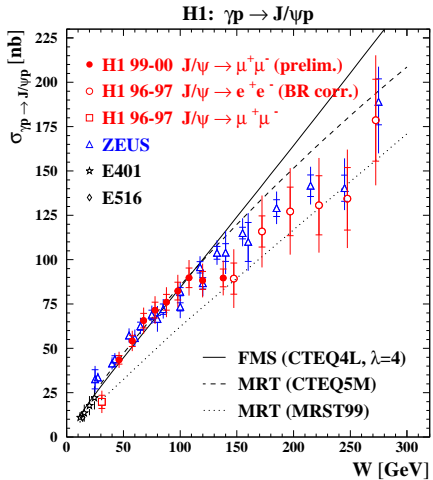}
\caption{The 
$J/\psi$
 cross sections 
 as a function
  of $W$. The experimental data are compared with pQCD models.
}
\label{fig:h1jpsi}
\end{figure}

The dependence of the cross section on $W$ is shown in
Fig.~\ref{fig:wdepxsect} and \ref{fig:h1jpsi} for exclusive vector
meson production 
along with the total cross section. The $W$ dependence
 for the lighter vector mesons ($\rho$, $\omega$, $\phi$) is similar to
that of the total cross section, as expected
from Regge theory.
However there is a clear change
as the vector meson mass increases: the $J/\psi$ cross section
has a steeper rise with $W$, which is a signature of a hard process.
In pQCD models, the cross section is proportional to the square of the
gluon density
of the proton
and
 the $W$ dependence
is then attributed to the 
rapid rise of the gluon density
at small $x$ (small $x\sim q^2/W$).

\begin{figure}
\epsfxsize180pt
\figurebox{120pt}{160pt}{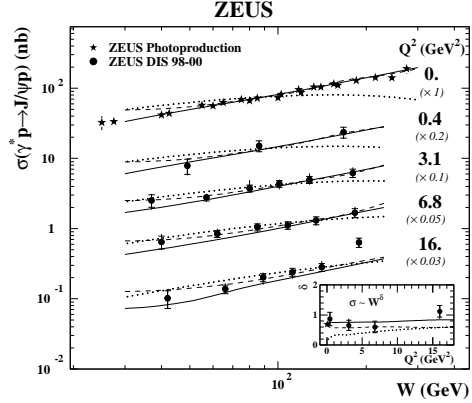}
\caption{Exclusive $J/\psi$ electroproduction cross section as a
  function of $W$ for four values of $Q^2$. The photoproduction
  results are also shown. The data are compared to MRT predictions
  obtained with different parametrisations of the gluon density
  (ZEUS-S solid line, CTEQ6M dashed
and MRST02 dotted)
 and
  normalised to the photoproduction data at $<W>=90~\mbox{GeV}$.
  The insert shows the parameter $\delta$ as a function of $Q^2$.
}
\label{fig:jpsipartons}
\end{figure}

The DIS $J/\psi$ data~\cite{jpsi,jpsih1}, shown  in
Fig.~\ref{fig:jpsipartons}, are precise enough to distinguish between
different parton density function sets, but no discrimination can be
made at present due to large theoretical uncertainties arising from
higher-twist corrections, missing higher-order terms and skewed parton
distributions. The data show that $W$ dependence of the $J/\psi$ cross section does
not change with $Q^2$~\cite{jpsidis}
and are well described by pQCD calculations~\cite{jpsitheo}.
This kind of behaviour is not followed in the case of the $\phi$ meson
where the $W$ slope increases slightly with $Q^2$~\cite{phi} as it is for
$\rho$ mesons~\cite{rho1,rho2}.

\begin{figure}
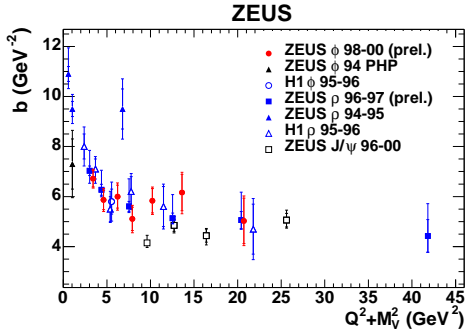

\epsfxsize180pt
\figurebox{120pt}{160pt}{bslope-rebinned-port.epsi}
\caption{The $b$ slope results of fits as described in the text, as a
function of $Q^2+M_V^2$, compared to previous H1 and ZEUS results. }
\label{fig:phi}
\end{figure}

Measurements of the
differential cross section, $d\sigma^{\gamma^* p\rightarrow\phi
p}/dt$, measured as a function of $t$ in the range
$|t|<0.6~\mbox{GeV}^2$ has also shown a variation of the slope for
different values of $Q^2$.  A function of the form
$d\sigma/dt=d\sigma/dt|_{t=0}\cdot e^{-b|t|}$ was fitted to the data
and the results are shown in Fig.~\ref{fig:phi}.  The results are
consistent with a global vector meson production scaling with
$Q^2+M_V^2$.

\section{Inelastic $J/\psi$ production}\label{sec:ineljpsi}

In the case of inelastic $J/\psi$ production another quantity is also
important: it is the inelasticity $z$ that, in the proton rest frame, is
equal to 
the fraction of the virtual photon energy taken by the $J/\psi$.

The direct and resolved photon processes can be described by the
colour singlet (CS) or colour octet (CO) approach. In the former,
 the $c\bar{c}$ pair is formed in a colour singlet state with
the $J/\psi$ quantum numbers and is directly identified with the $J/\psi$ physical
state. In the CO approach, the $c\bar{c}$ pairs are produced in colour
octet states which then evolve into $J/\psi$ mesons via the radiation of
soft gluons.  The transition of the
 $c\bar{c}$ pair into the physical
state is described in terms of long distance matrix elements tuned to
$J/\psi$ experimental data.

\begin{figure}
\epsfxsize180pt
\epsfysize170pt
\figurebox{120pt}{180pt}{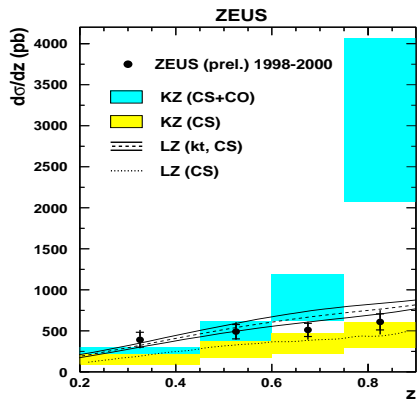}
\caption{The plot shows the inelasticity differential cross section
  measured by the ZEUS experiment for $Q^2>2~\mbox{GeV}^2$. The data
  are compared to the theoretical predictions described in the text.}
\label{fig:ineljpsi}
\end{figure}

Inelastic $J/\psi$ production in deep
inelastic scattering regime probes non-relativistic QCD
 at a higher scale.
Moreover
for $Q^2>0$ the resolved
photon processes, seen in photoproduction, are suppressed. Hence, in the DIS
regime, no enhancement of the cross
section at low $z$ is expected.

The ZEUS measurement of the differential cross section 
with respect to $z$ in the DIS
regime
\cite{ineljpsi}
 is shown in Fig.~\ref{fig:ineljpsi}. The
 data points are compared to a LO 
including both CS and CO or CS only matrix elements,
identified by the labels KZ (CS+CO) and KZ (CS),  respectively. As in
the photoproduction
regime~\cite{phprodjpsiinel1,phprodjpsiinel2}, the
strong rise of the CS+CO predicted cross section for $z>0.75$
  is not supported by data.  A similar observation in the DIS regime was
already reported by H1~\cite{disprodjpsiinel3}.  Predictions based on
the $k_T$ factorisation approach, identified by the label LZ(kt, CS),
are also shown. In this model, based on non-collinear parton dynamics
governed by the BFKL evolution equations, effects due to the non-zero
gluon transverse momentum are taken into account.
 The
cross section is calculated as the convolution of an unintegrated
 gluon density~\cite{unintgluon} with
 CS off-shell matrix element.
These calculations give reasonable description of the data both in
normalization and shape.

\section{Photon light-cone wave functions}\label{sec:pholightcone}

The internal structure of  the photon may be described using the
light-cone wave function formalism (LCWF). The photon LCWF has both electromagnetic
and hadronic components
 and
 can be studied for real or virtual
photons. The electromagnetic component of the photon LCWF can be
calculated within QED, while the hadronic one is model dependent.

In this contribution also 
the exclusive diffractive electroproduction of non-resonant
$\pi^+\pi^-$ pairs has been reported. This is expected to be sensitive
to the $|q\bar{q}\!>$ component in the photon LCWF. In high energy
interactions in the rest frame of the target, the valence Fock
component dominates, while the other terms are suppressed according to
counting rules~\cite{lcwfcounting}. In forward scattering, the
momentum configuration of the interacting Fock state is preserved and
therefore the final state is expected to reflect the kinematics of the
initial state. 

\begin{figure}
\epsfxsize180pt
\figurebox{120pt}{160pt}{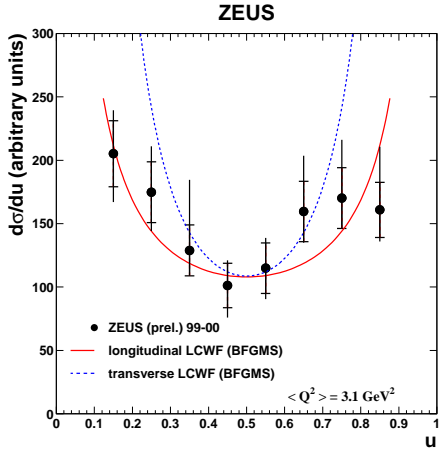}
\caption{Differential cross section $d\sigma/du$ measured for
  $40<W<120~\mbox{GeV}$, $1.2<M_{\pi\pi}<5~\mbox{GeV}$,
  $2<Q^2<5~\mbox{GeV}^2$ and $|t|<0.5~\mbox{GeV}^2$ .
    The data points are compared to the LCWF predictions, normalised to
  the data.
}
\label{fig:LCWF}
\end{figure}

Figure~\ref{fig:LCWF} shows the 
acceptance-corrected distribution
of the longitudinal-momentum fraction carried by one
of the pions, 
$(E_1+p_{z'_1})/(E_1+E_2+p_{z'_1}+p_{z'_2})$
where $E_1, E_2$ are the energies of the two pions and $p_{z'_1},
p_{z'_2}$
are the corresponding momentum components determined with respect to
the direction of the vector sum of the momenta of the two pions.
The results are compared with LCWF predictions 
 under the assumption that
the measured cross section is proportional to the square of the photon
wave function and that the pion momenta are equal to the momenta of the
initial $q\bar{q}$ pair.

The 
data are
compared to the
prediction for both transversely- and longitudinally-polarised
photons and are consistent with the 
latter.

\section*{Acknowledgements}
It is a pleasure to thank my colleagues in H1 and ZEUS for the
stimulating discussions. A special thank to Michele Arneodo for
the help in preparing this talk.

\end{document}